\documentclass[a4paper,twoside]{article}

\usepackage{epsfig}
\usepackage{subcaption}
\usepackage{calc}
\usepackage{amssymb}
\usepackage{amstext}
\usepackage{amsmath}
\usepackage{amsthm}
\usepackage{multicol}
\usepackage{pslatex}
\usepackage{apalike}

\usepackage{algorithm2e}
\usepackage[bottom]{footmisc}
\usepackage[nolist]{acronym}
\usepackage{dblfloatfix}
\usepackage{url} 
\usepackage{SCITEPRESS}     % Please add other packages that you may need BEFORE the SCITEPRESS.sty package.

\usepackage{epstopdf}

\SetAlCapHSkip{0cm}
\begin{document}

\title{Towards a Cloud-Based Smart Office Solution for Shared Workplace Individualization}

\author{\authorname{Dominik Hasiwar\sup{1}, Andreas Gruber\sup{1}, Christian Dragschitz\sup{1} and Igor Ivki\'c\sup{1,2}\orcidAuthor{0000-0003-3037-7813}}
    \affiliation{\sup{1}University of Applied Sciences Burgenland, Eisenstadt, Austria}
    \affiliation{\sup{2}Lancaster University, Lancaster, U.K.}
    \email{\{2210781002, 2210781009, 2210781021\}@fh-burgenland.at, i.ivkic@lancaster.ac.uk}
}

\keywords{Smart Office, Shared Workplace Individualization, Workplace Environment Index.}

\abstract{In the evolving landscape of workplace dynamics, the shift towards hybrid working models has highlighted inefficiencies in the use of traditional office space and the need for an improved employee experience.
	In this position paper we propose a Smart Office solution that addresses these challenges by integrating a microservice architecture with Internet of Things (IoT) technologies to provide a flexible, personalized workspace environment.
	The position paper focuses on the technical implementation of this solution, including the design of a Workplace Environment Index (WEI) to monitor and improve office conditions.
	By using cloud technology, IoT devices with sensors, and following a user-centred design, the proposed solution shows how Shared Open Workspaces can be transformed into adaptive, efficient environments that support the diverse needs of the modern workforce.
	This position paper paves the way for future experimentation in real-world office environments to validate the effectiveness of the Smart Office solution and provide insights into its potential to redefine the workplace for improved productivity and employee satisfaction.%\\\\\\\\.
}

\onecolumn \maketitle \normalsize \setcounter{footnote}{0}

\section{INTRODUCTION}
\label{sec:introduction}

The COVID-19 pandemic initiated a critical and unforeseen shift in workplace dynamics, forcing companies to quickly move from a traditional \textbf{\textit{zero home office}} policy to a \textbf{\textit{100\% home office}} model.
This drastic change out of necessity has gradually evolved into a hybrid work culture that includes a combination of remote and in-office work.
In the post-pandemic landscape, this shift has brought flexibility, improved employee satisfaction and significant reductions in operational costs such as heating, electricity, and maintenance.
However, it has also led to a noticeable inefficiency in the use of office space.
The changing presence of staff, with some working remotely while others are on site, often leaves office desks unoccupied.
This represents a clear inefficiency in the current office layout and highlights the need for more open and adaptable environments to optimize the use of office space and reflect the changing nature of work in the post-pandemic era \cite{Elias2023}.

Building on the idea of reconfigured office spaces, the concept of \textit{Shared Open Workspaces} offers a promising solution where, for example, employees can reserve desks for daily use.
This model increases the efficiency of office space use and fits well with the flexible nature of the hybrid working model. However, there is a notable trade-off in terms of personalization and comfort.
In such environments, the lack of customizable elements such as ergonomic adjustments, personal memorabilia (e.g. family pictures) and preferred desk locations can affect the sense of belonging and comfort at work.
In addition, the varying environmental conditions within the workspace, such as uneven temperature, humidity, or noise levels, can have a significant impact on an employee's productivity and well-being.
Some may find themselves in less than ideal conditions, which can lead to long-term discomfort or dissatisfaction.

Addressing these challenges requires an approach that improves the efficiency of \textit{Shared Open Workspaces} while allowing employees to personalize their individual workspace.
The overall aim of this approach should be to reconcile the benefits of office space optimization with the personal needs and comfort of each employee.

In response to the challenges identified, this position paper presents a cloud-native solution based on a microservice architecture, comprising three main services to improve the experience within \textit{Shared Open Workspaces}.
The first service (\textit{Workspace Management Service}) can be used to create and manage workspaces (offices) and their workplaces (desks), while the second service (\textit{Workplace Individualization Service}) enables workplace customization.
This includes personalizing of each workplace by enabling ergonomic adjustments (e.g., desk height) and personalized memorabilia (e.g., digital frames displaying private photographs).
This service also collects environmental data using sensors to monitor conditions such as temperature, humidity, and noise levels.
Finally, the third service (\textit{Workplace Booking Service}) offers employees the flexibility to reserve and activate their workspaces, which adapt to their preferences when activated.
Together, these services aim to create a dynamic, user-friendly, and adaptable \textit{Shared Open Workspace} environment.

The remainder of this paper is organized as follows: Section 2 summarizes the related work in the field. Next, in Section 3, we present the architectural building blocks of the proposed \textbf{\textit{Smart Office}} solution.
Furthermore, we propose a methodological approach to calculate a Workplace Environment Index (WEI) based on the collected environmental data.
In this context, we explain how the WEI can be used to rank workplaces, or even to improve their environmental conditions of them.
Finally, Section 5 concludes our work and outlines future work in the field.

\section{RELATED WORK}
\label{sec:relatedwork}

In the research area of smart offices, office productivity, and smart workplace environments, a number of studies have focused on the impact of various physical and technological factors on employee performance and well-being. Mak and Lui (2011)\nocite{Mak2011}, and Leaman (1995)\nocite{Leaman1995} evaluated the impact of environmental influences on productivity, highlighting the effects of sound, temperature, and overall satisfaction with workspace conditions.

Porras-Salazar et al. (2021)\nocite{PorrasSalazar2021} conducted a meta-analysis on the effects of room temperature on work performance in the office and examined 35 studies to assess the relationship between air temperature and cognitive performance in office environments.
Lusa et al. (2019)\nocite{Lusa2019} focused on multi-space offices, linking workspace design elements like furniture and acoustics to employee well-being and satisfaction.

Öhrn et al. (2021)\nocite{Ohrn2021}, and Hanif and Saleem (2020)\nocite{Hanif2020} examined the effects of office design variations, including Activity-Based Flex Offices and different university library layouts, on productivity and employee satisfaction. Pitchford et al. (2020)\nocite{Pitchforth2020} contributed an experimental perspective by comparing different office designs in a technology company, highlighting preferences for zoned open-plan and team office designs.

Robertson et al. (2013)\nocite{Robertson2013} focused on ergonomics, demonstrating the benefits of ergonomic training and sit-stand workstations on worker performance and discomfort. Wells' (2000)\nocite{Wells2000} study explored the role of personalization in the office, demonstrating its significant impact on employee well-being and job satisfaction, with gender-based differences in personalization practices.

In the research area of smart office environments, Zhang et al. (2022)\nocite{Zhang2022} reviewed Internet of Things (IoT) and Artificial Intelligence (AI) applications for employee health promotion, while Choi et al. (2015)\nocite{Choi2015} presented a smart office energy management system using Bluetooth Low Energy (BLE) beacons and a mobile app. Bhuyar and Ansari (2016)\nocite{Bhuyar2016} describe an integrated smart office automation system, demonstrating the usefulness of IoT technologies in improving office environments.

Ryu et al. (2015)\nocite{Ryu2015} proposed an Integrated Semantics Service Platform for smart offices, emphasizing semantic interoperability in IoT-based services. Tuzcuo\v{g}lu et al. (2015)\nocite{Tuzcuolu2022} provided a user-centric exploration of smart office environments, highlighting the importance of understanding user expectations. Finally, Uppal et al. (2021)\nocite{Uppal2021} introduced a cloud-based IoT system for smart offices, focusing on device health monitoring and fault prediction to enhance employee well-being.

Summarizing, the identified studies provide a multifaceted view of how physical, ergonomic, and technological elements interact to shape the modern office environment and influence employee productivity, satisfaction, and well-being. However, they have not fully explored the use of IoT devices combined with a management application to personalize workspaces and monitor environmental factors in a flexible workplace culture. The proposed \textbf{\textit{Smart Office}} approach aims to fill this gap by providing a comprehensive solution that combines workspace personalization with environmental monitoring and control, leveraging IoT and cloud technology.

\section{SMART OFFICE SOLUTION}

%-------------------------------------------------------------(Inserted by INSTICC:Fx)
\begin{figure*}[!b]\vspace{-0.25cm}
	\centering
	\includegraphics[width=\textwidth]{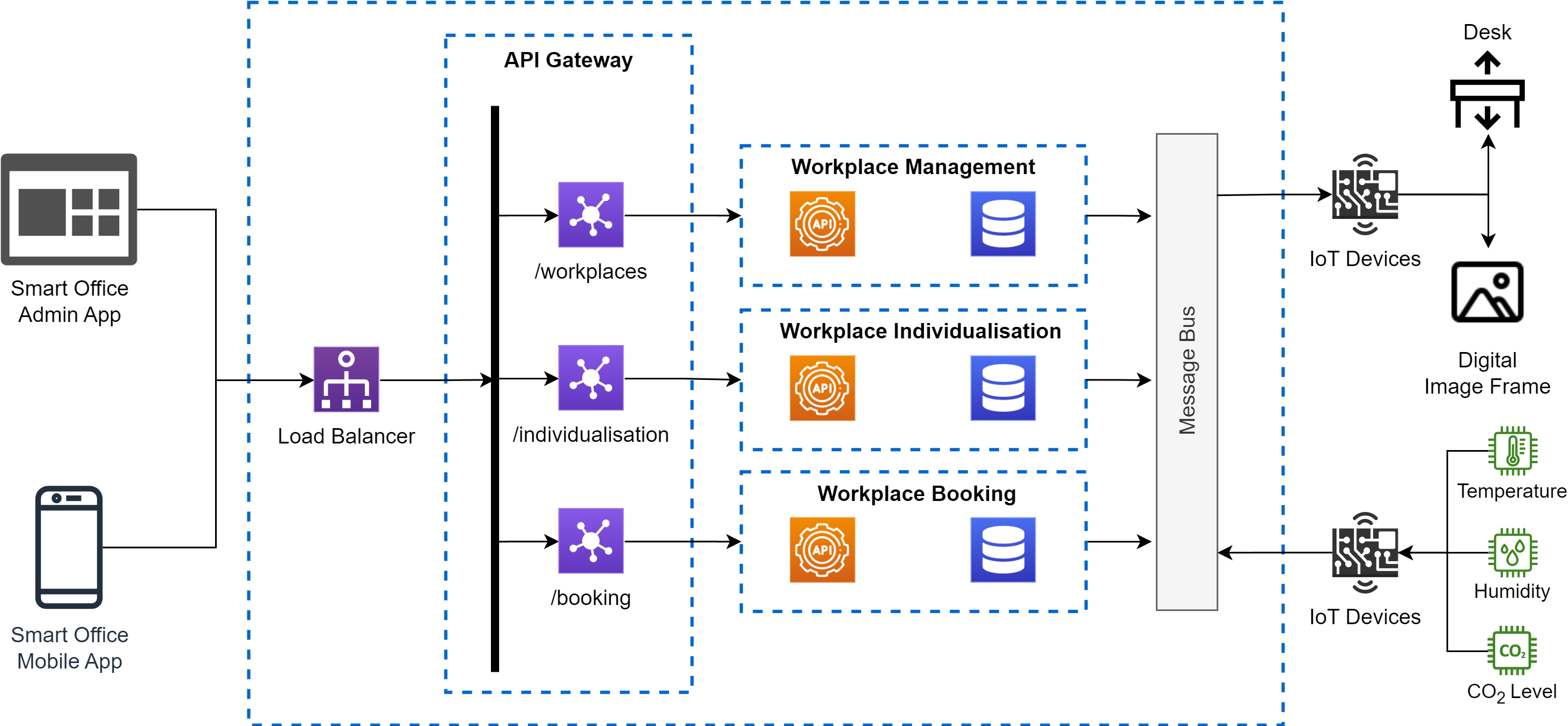}
\caption{Architectural Building Blocks of the Smart Office Solution.}
\label{fig:SmartOfficeMicroservices}
\vspace{0.25cm}
\end{figure*}

In this section, we present the technical implementation of the proposed \textbf{\textit{Smart Office}} solution. First, we provide an overview of the architectural building blocks and explain the high-level architectural design.
Next, we describe the three core microservices of the proposed solution and explain how they can be used to manage, individualize and book workplaces in a \textit{Shared Open Workspace} environment.
Finally, we present two mobile apps and explain how they can be used by regular users and administrators to interact with the \textbf{\textit{Smart Office}} solution.

\subsection{Architectural Building Blocks}

The \textbf{\textit{Smart Office}} is designed as a set of independent microservices, each providing a distinct functionality, yet seamlessly integrating to form a robust and agile application environment.
The solution also relies on a number of other components and managed services that are used in a typical microservice architecture.
Figure \ref{fig:SmartOfficeMicroservices} shows the individual components of the \textbf{\textit{Smart Office}} application environment.
This cloud-native approach not only improves the maintainability of the application, but also unlocks the full potential of cloud capabilities (managed services, auto-scaling, fault tolerance, and distributed processing).

At its core, the \textbf{\textit{Smart Office}} is orchestrated within Kubernetes, an open-source platform that automates the deployment, scaling, and management of containerized applications.
This ensures that each microservice is efficiently managed and scaled to meet the dynamic needs of the application requirements.
Kubernetes plays a central role in maintaining the resilience and elasticity of the application, enabling the seamless handling of variable workloads \cite{Sharma2022}.

Complementing the microservice architecture, the \textbf{\textit{Smart Office}} uses the database-per-service pattern, where each microservice has its own database.
This approach emphasizes the principle of loose coupling, by ensuring that each service is completely independent in terms of data management \cite{Kumar2017}.
These dedicated database instances are provided by a managed database service, which abstracts away the complexities of database management such as provisioning, scaling, backup, and recovery.
This ensures that the data processing aspect of the \textbf{\textit{Smart Office}} is robust, secure, and performance optimized.

In addition to the architectural components already mentioned, another important element of the \textbf{\textit{Smart Office}} is the message broker.
This application unit acts as an intermediary between the individual microservices and IoT devices, enabling the exchange of information.
Using a publish-subscribe model, the microservices and IoT devices can publish messages or subscribe to specific topics without having direct knowledge of each other.
This model increases the scalability and flexibility of the system, allowing services and devices to be easily added or changed without disrupting the entire application.

In addition to the \textbf{\textit{Smart Office}} microservices, the individual IoT devices installed on site at each workplace play an essential role in the overall functionality of the application.
These devices have two main functions. The first is to manage the customizable features of a workplace, such as operating a digital picture frame or controlling a height-adjustable desk. This functionality allows for a personalized and adaptable work environment, tailored to a user's individual preferences. The second function of the IoT devices is to monitor environmental conditions. Through connected sensors, they collect environmental data such as temperature, $CO_2$ concentration, and humidity to ensure a healthy and conducive working environment.

Although each of the components presented so far plays an important role in the overall application architecture, the \textbf{\textit{Smart Office}} microservices and mobile applications are at the heart of the application. The entire business logic of the solution is encapsulated in these application elements. The following sections provide a detailed introduction to their specific functions and responsibilities.

\subsection{Workspace Management Service}

By mapping each individual workplace as a digital object within the application, the Workspace Management Service provides a detailed and dynamic view of the office space and facilitates the efficient management and allocation of workplaces.
Beyond simple mapping, the Workspace Management Service carefully monitors the allocation of available workplaces, ensuring optimal utilization of the office space while also keeping track of the customizable attributes of each workplace.
These attributes, which currently include features such as a digital picture frame or a height-adjustable desk, greatly enhance the user experience by allowing the work environment to be tailored to individual preferences.
Importantly, the architecture of the system is designed for future adaptability, allowing additional features to be seamlessly integrated as new technologies emerge.

\subsection{Workplace Individualization Service}

The Workplace Individualization Service specializes in the management and aggregation of personalized data specifically tailored to individual work environments. By focusing on personalized data collection, this service plays an important role in optimizing the work experience through a tailored approach that addresses the unique needs of each workplace. The service's responsibilities fall into three distinct categories:
\begin{itemize}
	\item Ergonomic adaptability
	\item Personal memorabilia
	\item Environmental data collection
\end{itemize}

\noindent The first category focuses on the management of the data required to operate the height adjustable desks.
This aspect ensures that each workspace can be physically adapted to meet the ergonomic needs of its user, promoting health and comfort in the office workspace.
The second category relates to the management of personal memorabilia.
This involves managing users' pictures, which are displayed in digital picture frames and add a personal touch to any workplace.
In this way, the service helps to create a more familiar and welcoming working environment, boosting morale and a sense of belonging.
The final category relates to the environmental data collected by the IoT devices, such as temperature, CO\textsubscript{2} concentrations and humidity.
This information is then used to create the WEI which gives administrators a quick overview of the environmental status of each workplace.

\subsection{Workplace Booking Service}

The Workplace Booking Service is designed to allow users to reserve a specific (open) workplace for a specific period of time. In addition, the service provides a comprehensive view of workplace availability, ensuring that users can easily find and secure a space that meets their needs, thereby increasing the flexibility and efficiency of office space utilization.

As well as managing reservations, the Workplace Booking Service is also responsible for the actual activation of a booking.
Once a user has booked a workspace, the service works with other components of the \textbf{\textit{Smart Office}} application to personalize the workspace according to the user's preferences.
This personalization includes adjusting the height of the desk and displaying user-specific images on a digital picture frame.
By integrating information from multiple sources, the service not only simplifies the reservation process but also ensures that each workspace is tailored to provide a comfortable, personalized, and productive environment for each user.

\subsection{Mobile Applications}

The \textbf{\textit{Smart Office}} microservice application has been complemented by two specialized mobile applications, each designed to address different user requirements.
The first application is optimized for the regular users of the \textbf{\textit{Smart Office}}, offering a user-friendly interface and a range of features designed to enhance their daily office experience.
This includes the ability to make workplaces reservations, manage personal preferences and customize individual workplace configurations.
Users can personalize their office environment, from configuring their preferred images on digital frames to adjusting the height of their desks.

The second application is designed for administrators of the \textbf{\textit{Smart Office}} solution, enabling them to manage the complex aspects of office space utilization and maintenance.
Through the application, administrators can oversee the allocation of workspaces, consolidate user bookings, and monitor the overall usage of office resources.
In addition, the app provides access to a wealth of environmental data collected from various IoT sensors, including the insightful WEI.
This feature enables administrators to make informed decisions about workspace management, ensuring optimal use of office space and maintaining a high standard of workplace environment.

\section{WORKPLACE ENVIRONMENT INDEX (WEI)}

The WEI represents a significant step forward in the management and optimization of workplace conditions, harnessing the power of IoT technology.
The index is calculated using data collected by an IoT device using various sensors installed in a workplace.
These sensors measure various environmental parameters such as temperature, humidity and $CO_2$ concentration.
By continuously collecting and analysing this data, the WEI provides a comprehensive, real-time assessment of a workplace's environmental conditions.
The calculation of the WEI is based on the following general equation:

\vspace{-0.25cm}
\begin{equation}
    WEI=\frac{WEI_{1}*W_{1} + WEI_{2}*W_{2}+ ... + WEI_{n}*W_{n}}{n}
\end{equation}

As shown in (1), the WEI is calculated by the sum of weighted ($W_1$, $W_2$, ..., $W_n$) workplace environment measurements ($WEI_1$, $WEI_2$, ..., $WEI_n$) using the IoT device and its sensors divided by the total number of WEI measurements ($n$). The resulting measurement of each sensor represents a specific WEI (e.g., $WEI_1$) with a corresponding weight (e.g., $W_1$) to be able to emphasize one environmental measurement over another one.
This allows to prioritize certain WEI measurements and give some values more weight compared to others.
For instance, a certain workplace environment measurement (e.g., temperature) might be more important for a certain workplace compared to another one (e.g., humidity).
In this case the weight of the first measurement would be higher compared to the second to show which one is more significant in this specific case.

The WEI in (1) enables to calculate the environment index of a workplace to evaluate its environmental conditions.
This index can further be used when analysing the booking pattern of certain workplaces to identify the reasons why some are overbooked while others are not favoured by the employees.
In the following sections, the WEI calculation will be explored in more detail.

\subsection{WEI Calculation}

Based on the general WEI equation from (1), the calculation of the WEI follows a four-stage approach to ensure accuracy and relevance.

\vfill

\subsubsection{Step 1: Define Range of Values}

The first step is to establish a range of values for the environmental data to be measured by determining an optimum, minimum and maximum value for each environmental parameter. This first step is essential as it lays the foundation for the subsequent normalization process to ensure that the data accurately reflects the quality of the workplace environment. The following table shows an example range of values for the environmental values of temperature, humidity and CO\textsubscript{2} concentration:

%-------------------------------------------------------------(Inserted by INSTICC:Fx)
\begin{table}[ht]
\vspace{0.2cm}
\caption{Min-max Value Range of Environmental Data.}
\label{tbl:minMaxValueRange}

    \small
    \centering
    \begin{tabular}{ |p{2cm}|p{2cm}|p{2cm}|  }
        \hline
                                  & Minimum        & Maximum        \\
        \hline
        Temperature               & 15 $^{\circ}$C & 30 $^{\circ}$C \\
        \hline
        Humidity                  & 10\%           & 80\%           \\
        \hline
        CO\textsubscript{2} Level & 100 ppm        & 800 ppm        \\
        \hline
    \end{tabular}
\vspace{-0.3cm}
\end{table}

\subsubsection{Step 2: Normalization}

Once the value ranges are established, the second step is to apply a linear min-max normalization, as proposed by Vafaei et al. (2016)\nocite{Vafaei2016} using the following equation:

\vspace{-0.15cm}
\begin{equation}
    n_{ij}=\frac{r_{ij}-r_{min}}{r_{max}-r_{min}}
\end{equation}

The min-max normalization from (2) converts the measured environmental values into a normalized range from 0 to 1, where the value 0.5 represents the previously defined optimum or mean value for each parameter. This normalized scale ensures that the mean value accurately reflects the most desirable environmental condition and allows a consistent and fair assessment of different environmental parameters. Based on this, the environmental measurements of different metrics (with different units) can be aggregated after the normalization step, allowing the calculation of an overall WEI as shown in (1).

\subsubsection{Step 3: Transformation}

In the next step, the normalized values are converted to the actual WEI value range of 0 to 100. This conversion is based on the use of the following parabolic function:

\vspace{-0.25cm}
\begin{equation}
    f(x)=a(x-h)^2+k
\end{equation}

The idea is to have a parabola that peaks at 0.5 and intersects the x-axis at 0 and 1, corresponding to the minimum and maximum of the normalized values. The variables \textit{h} and \textit{k} represent the vertex coordinates of the parabola. As a result, the variable \textit{h} represents the previously established optimal normalized value of 0.5, while \textit{k} represents the upper limit of the WEI scale, which is 100. Solving the equation for the points of intersection of the parabola with the x-axis gives the actual value of the variable \textit{a} \cite{Papula2018}. Table \ref{tbl:weiCalculation} shows the conversion of the measured example environmental data into the WEI using the parabolic function:

%-------------------------------------------------------------(Inserted by INSTICC:Fx)
\begin{table}[ht]
\vspace{0.2cm}
\caption{Normalized environmental data.}
\label{tbl:weiCalculation}
\centering
    \small

    \begin{tabular}{ |p{1.6cm}|p{1.4cm}|p{1.5cm}|p{1.2cm}|  }
        \hline
                                  & Measured        & Normalized & WEI   \\
        \hline
        Temperature               & 23.5$^{\circ}$C & 0.57       & 98.04 \\
        \hline
        Humidity                  & 35\%            & 0.36       & 92.16 \\
        \hline
        CO\textsubscript{2} Level & 600 ppm         & 0.71       & 82.36 \\
        \hline
    \end{tabular}
\vspace{-0.2cm}
\end{table}

As shown in Table \ref{tbl:weiCalculation}, the initial measured temperature value of 23.5°C was normalized using the min-max normalization from (2) with its minimum and maximum values from Table \ref{tbl:minMaxValueRange}. Then the parabolic function from (3) was used to convert the normalized value of 0.57 to the WEI of 98.04.

\subsubsection{Step 4: Aggregation}

Once all measured values have been converted, the final step is to aggregate them to calculate the overall WEI for a specific workplace. As previously mentioned, there may be cases where not all the calculated values contribute equally to the overall WEI. For this reason, weights are used to emphasize one result or environmental measure over another. This means that specific environmental factors can be given more weight than others, depending on their impact on the workplace environment. Based on the weighted approach, the final score provides a tailored assessment of the workplace environment that reflects the unique priorities and needs of each workplace. As a result, the partial WEI results for the temperature, humidity and the CO\textsubscript{2} level as well as the strategic weighting of the environmental variables are incorporated into the subsequent equation:

\vspace{-0.25cm}
\begin{equation}
    WEI=\frac{WEI_{T}*W_{T} + WEI_{H}*W_{H} + WEI_{C}*W_{C}}{3}
\end{equation}

Using the example environmental measurement data from Table \ref{tbl:minMaxValueRange} and giving equal weight to each metric, the overall WEI result is 90.85.

\subsection{WEI Applicability}

The calculated WEI can be used as an indicator to evaluate different environmental conditions of a workplace and calculate an overall score. This approach enables to monitor the conditions of a workplace in real-time and to take action to improve or optimize the conditions of a workplace.
In addition, the use of IoT technology demonstrates how a physical work environment can become smart by dynamically adapting to the user needs.

One of the key advantages of the WEI is its utility in enhancing user experience within a shared workspace environment.
Employees within the \textbf{\textit{Smart Office}} can select workplaces that perfectly match their personal preferences and comfort requirements based on the calculated WEI.
This capability is not only beneficial for satisfying employees' comfort needs, but also for optimizing their productivity and well-being.
In addition, the WEI can be used to rank workplaces, providing a clear, data-driven guide to the best performing environments within a facility.
Such a feature democratizes workplace choice and ensures that all users have access to the spaces that best suit their needs, thereby promoting a more equitable and satisfying working environment.

Beyond the benefits to the individual user, the WEI can be an invaluable tool for facilities management, acting as an early warning system to highlight potential problems at specific workstations.
For example, abnormal temperature readings could indicate a malfunction in the ventilation system, while elevated CO\textsubscript{2} levels could indicate overcrowding in a particular area.
Furthermore, by correlating WEI data with user booking patterns, facility managers can gain insight into the environmental conditions that are most conducive to employee comfort and productivity.

This improved understanding can contribute to more efficient control of facility systems for heating, cooling and lighting based on real-time occupancy and usage patterns.
Such tailored environmental adjustments not only improve the overall workplace experience but also contribute to energy efficiency and sustainability goals.
In the area of predictive analytics, trends in WEI data can predict the need for maintenance or indicate potential failures in facility services, enabling pre-emptive action to minimize disruption and improve overall workplace functionality.

\section{CONCLUSIONS AND FUTURE WORK}
\label{sec:conclusion}

The COVID-19 pandemic has been the driving force behind an unprecedented shift towards hybrid and remote working models, opening up a wide range of possibilities for reimagining the traditional office space. At the heart of this transformation is the challenge of optimizing these new working environments to meet the diverse needs of a dispersed workforce, while maximizing efficiency and fostering a sense of belonging among employees.
In this paper we proposed a cloud-based \textbf{\textit{Smart Office}} solution for \textit{Shared Open Workspaces}, with a focus on individualization, efficiency within adaptive work environment.

First, we presented a technological framework using microservice architecture to manage, personalize, and book office spaces, complemented by mobile applications for end users and administrators. Next, we presented a model that uses different environmental measurement metrics (measured by an IoT device with sensors) and combines them to calculate an overall WEI score. Finally, we discussed how the calculated WEI can be used to monitor and optimize workplace conditions, improving employee well-being, and productivity.

The proposed  \textbf{\textit{Smart Office}} solution bridges the gap between workspace flexibility and personalization needs in hybrid working models. Based on a microservice architecture and IoT integration, the solution enables scalable, user-centric office environments. In addition, the WEI score provides actionable insights into workplace conditions, promoting optimal workspace utilization and employee satisfaction.

To further support the proposed \textbf{\textit{Smart Office}} solution, it is essential to conduct empirical studies or pilot implementations.
Therefore, it is planned to implement the \textbf{\textit{Smart Office}} solution in a real-world office environment, incorporating adjustable desks and digital frames for a personalized touch.
This empirical real-world implementation serves as a foundational step not only for assessing the solution's effectiveness and practicality but also for analysing critical aspects such as cost efficiency, security, privacy concerns, and scalability.
By evaluating these facets, the implementation aims to ensure the solution's viability in diverse settings and its adaptability to various user needs and office environments.

The empirical approach will be complemented by a user experience study to correlate user satisfaction with the WEI outcomes, thereby enabling iterative refinement of the solution based on actual usage data.
A thorough evaluation of the WEI in a live environment will help to identify and improve both optimal and underperforming workspaces.
The adaptability of the \textbf{\textit{Smart Office}} solution to offices of varying sizes and organizational structures is a key consideration.
A detailed scalability study will be conducted to evaluate how the solution can be effectively scaled up or down, depending on the specific needs of an organization.
This will include exploring modular aspects of the microservice architecture and the IoT framework to ensure that the \textbf{\textit{Smart Office}} solution can be tailored to accommodate different workspace environments and employee densities.

This position paper paves the way for future research into  \textbf{\textit{Smart Office}} environments, focusing on sustainability, user engagement, and the integration of advanced technologies to further enhance the office of the future.

\renewcommand{\refname}{REFERENCES}  % (Automatically inserted by INSTICC:Fx)

\bibliographystyle{apalike}
{\small\bibliography{references}}

\vfill

\end{document}